\documentstyle[prd,aps,twocolumn]{revtex}

\tighten

\begin{document}
\draft
 
\pagestyle{empty}

\preprint{
\noindent
\hfill
\begin{minipage}[t]{3in}
\begin{flushright}
LBNL--49672 \\
UCB--PTH--02/09 \\
hep-ph/0202222 \\
\end{flushright}
\end{minipage}
}

\title{Helicity conservation and 
factorization-suppressed charmless $B$ decays}

\author{
Mahiko Suzuki
}
\address{
Department of Physics and Lawrence Berkeley National Laboratory\\
University of California, Berkeley, California 94720
}


\date{\today}
\maketitle

\begin{abstract}

Toward the goal of extracting the weak angle $\alpha$, the 
decay $B^0/\overline{B}^0\to a_0^{\pm}\pi^{\mp}$ was recently measured.
The decay $B^0\to a^+\pi^-$ is not only forbidden in the factorization 
limit of the tree interaction, but also strongly suppressed for 
the penguin interaction if short-distance QCD dominates. This makes
extraction of $\alpha$ very difficult from $a_0^{\pm}\pi^{\mp}$. 
We examine the similar factorization-suppressed decays, in particular, 
$B^0\to b_1^+\pi^-$. The prospect of obtaining $\alpha$ is even less 
promising with $b_1^{\pm}\pi^{\mp}$. To probe how well the 
short-distance dominance works, we emphasize importance of testing 
helicity conservation in the charmless $B$ decays with spins.  

\end{abstract}
\pagestyle{plain}
\narrowtext

\setcounter{footnote}{0}

\section{Introduction}

  The BaBar Collaboration recently measured the branching fraction
for $B^0/\overline{B}^0\to a_0^{\pm}\pi^{\mp}$ to be 
$(6.2_{-2.5}^{+3.0}\pm 1.1)\times 10^{-6}$\cite{BaBar}.  
This is comparable to ${\rm B}(B^0\to\pi^+\pi^-)\simeq 4.4\times 10
^{-6}$\cite{CR}. The tree interaction does not contribute to 
$B^0\to a_0^+\pi^-$ in the factorization limit 
since $a_0^+$ cannot be produced from the $V-A$ current\cite{Laplace}. 
Contrary to what was alluded in Ref. \cite{Laplace}, however, 
this decay should be strongly suppressed even for the penguin 
interaction if the perturbative QCD picture is correct. Therefore 
the observed branching fraction should consist almost entirely of 
$B^0\to a_0^-\pi^+$ and $\overline{B}^0\to a_0^+\pi^-$. Consequently 
the time-dependent $B^0/\overline{B}^0\to a_0^{\pm}\pi^{\mp}$ 
decay will not be suitable for extraction of the angle $\alpha$ 
since there is little $B^0$-$\overline{B}^0$ interference 
in these channels\cite{Laplace,DK}. 

  We shall first show that the $B^0\to a_0^+\pi^-$ decay amplitude 
is power suppressed by $1/m_B$ for the penguin interaction if 
short-distances dominate. A numerical estimate will be given for the 
decay amplitude. We then examine other factorization-suppressed 
decays, in particular, production of $b_1^+$ that shares the same 
chiral property with the $a_0^+$ production. The chance of measuring 
the $B^0$-$\overline{B}^0$ interference is even slimmer for 
$b_1^{\pm}\pi^{\mp}$. The underlying assumption leading to these 
conclusions is that perturbative QCD is valid even for
the final-state interactions below $m_B$, {\em i.e.,} the
perturbative-QCD-improved factorization\cite{B,L}.  This assumption
need to be tested. For this purpose we discuss 
helicity conservation of light quarks in the $B$ decay, in general. 
It leads us to zero-helicity dominance in the charmless $B$ decay 
into two mesons both with spin. This selection rule provides us 
the simplest test of perturbative QCD in final-state interactions.
  
\section{Factorization-suppressed process}

\subsection{$B^0\to a_0^+\pi^-$}

We study for definiteness the decay amplitudes 
of the $B (\overline{b}q)$ meson instead of the $\overline{B}$ meson.
The tree interaction ${\cal O}_{1,2}$, {\em i.e.,} $(\overline{b_L}u_L)
(\overline{u_L}d_L)$ and its color-crossing, cannot produce $\pi^-$  
from the factorized current. It cannot produce $a_0^+$ either since 
G-parity does not match between $a_0^+$ and $\overline{u}\gamma^{\mu}d$. 
It appears therefore that the penguin interaction dominates in this
decay. If so, we would have an opportunity to extract the weak angle 
$\alpha$ from the time-dependent decay of 
$B^0/\overline{B}^0\to a_0^{\pm}\pi^{\mp}$\cite{Laplace,DK}. 
However, the penguin decay amplitude is suppressed by $1/m_B$ 
so that the $B^0$-$\overline{B}^0$ interference is too small 
for this purpose.

The QCD penguin operators, normally referred to as ${\cal O}_{5,6}$, 
generate the scalar density $\overline{u_L}d_R+\overline{u_R}u_L$ by
crossing. In the quark model its matrix element for the $a_0^+$ 
production is 
\begin{equation}
 \langle u({\bf k})\overline{d}(-{\bf k})|\overline{u}d|0\rangle =
 2{\bf k}\cdot\langle\mbox{\boldmath{$\sigma$}}\rangle \label{a0}
\end{equation}
where $\langle\mbox{\boldmath{$\sigma$}}\rangle =
\chi^{\dagger}\mbox{$\boldmath{\sigma}$}\chi'$ with $\chi$ and 
$\chi'$, the Pauli spinors of $u$ and $\overline{d}$, 
respectively. Superposing Eq. (\ref{a0}) with the wave function  
$\Phi({\bf k})$ of a $p$-wave bound state, we define the $a_0$ decay 
constant $f_{a0}$ by
\begin{equation}
   \langle a_0^+|\overline{u}d|0\rangle = f_{a_0}m_{a0}. \label{a02}
\end{equation}
In the $B^0$ rest frame where $u$ and $\overline{d}$ fly fast 
with total momentum ${\bf p}={\bf p}_u+{\bf p}_d$, 
\begin{equation}
   \langle (u\overline{d})({\bf p})|\overline{u}d|0\rangle =
 2{\bf k}_{\perp}\cdot\langle\mbox{\boldmath{$\sigma$}}_{\perp}
 \rangle + 2E_{{\bf k}}\beta(2x-1)\langle\sigma_{\parallel}\rangle,
                    \label{a03}
\end{equation}
where $\beta = |{\bf p}|/E_p$ ($E_p=\sqrt{m_{u\overline{d}}^2 +
{\bf p}^2}$) and $x=|{\bf p}_{u\parallel}|/|{\bf p}|$ with $\parallel$ 
and $\perp$ referring to the parallel and perpendicular direction 
to the momentum ${\bf p}$.
The right-hand side of Eq. (\ref{a02}) is $O(1)$ in $E_p$ in the
fast moving frame, which is consistent with the right-hand side of 
Eq. (\ref{a03}) after superposition with the light-cone distribution 
function $\Psi(x,{\bf k}_{\perp})$. It is one power lower in $E_p$ 
than, for instance, the $\pi^+$ production from 
$\overline{u}\gamma^{\mu}\gamma_5d$. This difference is due partly 
to the Lorentz property, but more importantly to the chiral property 
of quark fields involved, namely, $\overline{L}R\pm\overline{R}L$ 
{\em vs} $\overline{L}L\pm\overline{R}R$. 
We can perform the same calculation for the $^3P_1$ state, namely
$a_1$. The matrix element of quark-pair production  
$\langle u\overline{d}|\overline{u}\mbox{\boldmath{$\gamma$}}
\gamma_5d|0\rangle$ is equal to $2i\langle{\bf k}\times
\mbox{\boldmath{$\sigma$}}\rangle$ in the $a_1$ rest frame. 
If we superpose it with the same $p$-wave orbital wave function 
as $a_0$ and boost it to the $B^0$ rest frame, we obtain
\begin{equation}
   \langle a_1^+({\bf p})|\overline{u}\gamma^{\mu}\gamma_5d|0\rangle
         = f_{a1}m_{a1}\varepsilon^{\mu}({\bf p}),
\end{equation}
with $f_{a1}=\sqrt{2/3}f_{a0}$ in the constituent quark model. Since 
$f_{a1}\simeq f_{\rho}$\cite{Weinberg}, we find $f_{a0}\approx
f_{\rho}(\simeq f_{\pi})$.
A similar calculation for $a_2^+(^3P_2)$ 
leads us to $\langle a_2^+({\bf p})|\overline{u}\gamma^{\mu}
\stackrel{\leftrightarrow}{\partial^{\nu}}d|0\rangle = f_{a2}m_{a2}^2
\varepsilon^{\mu\nu}({\bf p})$ with $f_{a2}=\sqrt{1/2}f_{a0}$.
 
The formation of $\pi^-$ by ${\cal O}_{5,6}$ in $B^0\to a_0^+\pi^-$ 
is described by the scalar form factor of $B^0\to\pi^-$,
\begin{equation}
   \langle \pi^-|\overline{b}u|B^0\rangle = m_BF_s(q^2).
\end{equation}
The relation $i\partial_{\mu}(\overline{b}\gamma^{\mu}u)=(m_u-m_b)
(\overline{b}u)$ relates $F_s(q^2)$ to the vector form factors of
$B^0\to\pi^-$ as: $m_b m_B F_s(q^2)=(m_B^2-m_{\pi}^2)F_1(q^2) +
q^2F_2(q^2)$ so that $F_s(q^2)\simeq F_1(q^2)$ for $|q^2|\ll m_B^2$.

Expressing the penguin amplitude $A_p(B^0\to a_0^+\pi^-)$ in terms 
of the tree amplitude ($A_t$) of $B^0\to \pi^+\pi^-$, we have
\begin{eqnarray}
    \frac{A_p(B^0\to a_0^+\pi^-)}{A_t(B^0\to\pi^+\pi^-)} &\simeq&
    \frac{2V_{tb}V_{td}^*C_p(m_b)f_{a0}m_{a0}m_BF_s(m_{a0}^2)}{
    V_{ub}V_{ud}^*C_t(m_b)f_{\pi}m_B^2 F_1(m_{\pi}^2)},\nonumber\\
    &\simeq& 0.04\times [F_s(m_{a0}^2)/F_1(m_{\pi}^2)],
\label{a+}
\end{eqnarray}
where $C_t$ and $C_p$ are from the Wilson coefficients of the tree 
and penguin operators, respectively. The value $|V_{td}|\approx 0.01$
has easily 50\% uncertainty. Since the form factors scale as
$1/(1-q^2/m_B^2)$ with a relevant $B$-meson mass in the simple pole
approximation, we may set $F_s(m_{a0}^2)\simeq F_s(m_{\pi}^2)
\simeq F_1(m_{\pi}^2)$. Then the right-hand side is less than a tenth.
The main source of this suppression is the kinematical factor $1/m_B$
that is traced to the chiral suppression of $a_0^+$ production in
the fast moving frame.  

In comparison, the tree-allowed decay $B^0\to a_0^-\pi^+$ is
given for $f_{a0}\simeq f_{\pi}$ by
\begin{equation}
     \frac{A_t(B^0\to a_0^-\pi^+)}{A_t(B^0\to\pi^+\pi^-)}\simeq
 \frac{F_1^A(m_{\pi}^2)}{F_1(m_{\pi}^2)},
               \label{a-}
\end{equation}
where $F_1^A(q^2)$ is the axial-vector form factor $\propto 
(p_B+p_{a0})^{\mu}$ of $B^0\to a_0^-$. Since $F_1^A(m_{\pi}^2)
\approx F_1(m_{\pi}^2)$ is not far out of line, 
the right-hand side is about unity. Then 
Eqs. (\ref{a+}) and (\ref{a-}) are in line with the measured 
branching fraction\cite{BaBar} if it consists almost entirely of 
$B^0\to a_0^-\pi^+$ and $\overline{B}^0\to a_0^+\pi^-$. It means
little $B^0$-$\overline{B}^0$ interference in the $a_0^{\pm}\pi^{\mp}$ 
channels. 

For the decay ${\rm B}(B^0\to a_0^0\pi^0)$, all of ${\cal O}_{1,2}$, 
${\cal O}_{3,4}$, and ${\cal O}_{5,6}$ possibly contribute 
with comparable magnitudes so that their sum involves much a larger
uncertainty. Therefore a clean extraction of the weak angles
will be difficult from $B^0/\overline{B}^0\to a_0^0\pi^0$.  
It is obvious that Eqs. (\ref{a+}) and (\ref{a-}) apply to the 
ratios of $B^0\to a_0^{\pm}\rho^{\mp}$ to $B^0\to\pi^{\pm}\rho^{\mp}$
as well, if we replace the transition form factors appropriately. 
 
The QCD corrections turn the local operator $\overline{q}(x)\Gamma 
q(x)$ into the nonlocal operator $\overline{q}(x)q(y)$. But 
$\overline{q}(x)q(y)$ can be expanded in the Taylor series of local
operators in powers of $(x-y)_{\mu}\partial^{\mu}$. If 
short-distance interactions dominate, $|x-y|$ is a fraction of 
$1/m_b$ so that all terms of the Taylor expansion have the same
$1/m_B$ dependence as the leading term, {\em i.e.,} the local operator. 
Evaluation of the higher derivative terms require knowledge of more 
than a decay constant. When a meson cannot be produced from a local 
operator appearing in the effective interactions ${\cal O}_i$, 
it should be noted that the leading contribution is accompanied by
$\alpha_s/\pi$ due to a QCD loop. This is the case for $b_1$ 
and $a_2$. In such a case the amplitude of the leading order in 
$1/m_B$ depends on the shape of the distribution function of 
the $p$-wave bound states of which we know less.

If $a_0$ is a four-quark state $q\overline{q}q\overline{q}$ instead 
of a $q\overline{q}$ in $^3P_0$, we can show with a dimension argument 
of perturbative QCD that the $B^0\to a_0^{\pm}\pi^{\mp}$ decay 
amplitudes are even more suppressed in $1/m_B$ than 
$A_p(B^0\to a_0^+\pi^-)$. The positive identification of 
$B^0/\overline{B}^0\to a_0^{\pm}\pi^{\mp}$\cite{BaBar} is 
an evidence against the four-quark assignment of $a_0$ or 
else for breakdown of perturbative QCD.

\subsection{$B^0\to b_1^+\pi^-$}

Production of other $^3P_J$ states, $a_1^+\pi^-$ and $a_2^+\pi^-$, 
is different from that of $a_0^+$. They are produced from  
$(\overline{u_R}\gamma^{\mu}d_R)-(\overline{u_L}\gamma^{\mu}d_L)$ 
and $(\overline{u_R}\gamma^{\mu}
\stackrel{\leftrightarrow}{\partial^{\nu}}d_R) + 
(\overline{u_L}\gamma^{\mu}
\stackrel{\leftrightarrow}{\partial^{\nu}}d_L)$, respectively.
While the axial-vector current is found in the tree interaction, 
the tensor operator must be generated by gluon corrections. 
Therefore its derivative $\stackrel{\leftrightarrow}{\partial^{\nu}}$ 
comes with $\alpha_s/\pi$ and also with $1/E \simeq |x-y| = O(2/m_b)$. 
Since the tree interaction contributes in full strength to
production of $a_1^+$ and of $a_2^+$, these decays are simply
related to the tree-dominated $B^0\to\pi^+\pi^-$ decay as:
\begin{eqnarray}
 \frac{{\rm B}(B^0\to a_1^+\pi^-)}{{\rm B}(B^0\to\pi^+\pi^-)}&\simeq&
  \biggl|\frac{f_{a1}F_1(m_{a1}^2)}{f_{\pi}F_1(m_{\pi}^2)}\biggr|^2
                       \approx 1,                 \nonumber\\
 \frac{{\rm B}(B^0\to a_2^+\pi^-)}{{\rm B}(B^0\to\pi^+\pi^-)}&\approx&
 \biggl(\frac{\alpha_s(E)}{\pi}\biggr)^2 
 \biggl|\frac{f_{a2}E_{a2}F_1(m_{a2})}{f_{\pi}EF_1(m_{\pi})}\biggr|^2.
\end{eqnarray} 
While a considerable uncertainty exists in the relevant value of 
$\alpha_s$, we reasonably expect from the second line with 
$E\simeq E_{a2}$ that ${\rm B}(B^0\to a_2^+\pi^-)\leq 10^{-2}\times
{\rm B}(B^0\to\pi^+\pi^-)$. 

   Production of $b_1^+$ ($^1P_1$) has similarity with $a_0$ in the
chiral structure and with $a_2$ in the $\alpha_s/\pi$ suppression. 
The local operator that matches the quantum numbers of 
$b_1^+$ is $i(\overline{u}\gamma_5\stackrel{\leftrightarrow}{
\partial^{\mu}}d)$, whose chiral property is $\overline{L}R - 
\overline{R}L$. In the $b_1$ rest frame,
\begin{equation}
    \langle u({\bf k})\overline{d}(-{\bf k})|i\overline{u}\gamma_5
      \stackrel{\leftrightarrow}{\partial^{\mu}}d|0\rangle =
          2E_{{\bf k}}{\bf k}\langle 1\rangle, \label{b1}
\end{equation}
where $\langle 1\rangle = \chi^{\dagger} 1 \chi'$. Superposing Eq. 
(\ref{b1}) with the same $p$-wave orbital function as the $^3P_J$ 
mesons, we obtain
\begin{equation}
  \langle b_1({\bf p})|i\overline{u}\gamma_5
 \stackrel{\leftrightarrow}{\partial^{\mu}}d|0\rangle
      = f_{b1}m_{b1}^2\varepsilon^{\mu}(p) \label{b2}
\end{equation}
with $f_{b1}=\frac{1}{2}f_{a1}$ in the quark model. In
the $B^0$ rest frame,
\begin{equation}
    \langle u\overline{d}({\bf p})|\overline{u}\gamma_5
      \stackrel{\leftrightarrow}{\partial^{\mu}}d|0\rangle =
    \left\{  \begin{array}{ll}
 2\gamma\beta E_{{\bf k}}k_{\parallel}\langle 1\rangle,&(\mu=0) \\
 2\gamma E_{{\bf k}}k_{\parallel}\langle 1\rangle, & (\mu=\parallel)\\
 2E_{{\bf k}}{\bf k}_{\perp}\langle 1\rangle, & (\mu = \perp), 
        \end{array}  \right.  \label{b3}
\end{equation}
where $\gamma = (1-\beta^2)^{-1/2}$ so that $\gamma E_{{\bf k}}
\simeq  E_p$, the $b_1$ energy in the $B^0$ frame. 
The right-hand side of Eq. (\ref{b3}) is therefore $O(E_p)$ for
the time and longitudinal components in the fast moving frame, 
and $O(1)$ for the
transverse components, which is consistent with Eq. (\ref{b2}). 

As pointed out above, the derivative $\stackrel{\leftrightarrow}{
\partial^{\mu}}$ is accompanied by $\alpha_s(E)/\pi E$ so that it 
does not enhance the high-energy behavior when short distances 
dominate, {\em i.e.,} $E=O(\frac{1}{2}m_b)$. As in the case of 
$a_0^+$, the high-energy behavior of $\overline{L}R-\overline{R}L$ 
is lower by one power of energy in the fast moving frame.
Consequently the decay branching fraction ${\rm B}(B^0\to b_1^+\pi^-)$ 
scales just like ${\rm B}(B^0\to a_0^+\pi^-)$, namely $1/m_B^2$ down
relative to the allowed-tree branching fraction:
\begin{equation}
 \frac{{\rm B}(B^0\to b_1^+\pi^-)}{{\rm B}(B^0\to a_0^+\pi^-)}\approx 
 \biggl(\frac{\alpha_s(E)}{\pi}\biggr)^2
 \biggl|\frac{F_1(m_{b1}^2)}{F_1(m_{a0}^2)}\biggr|^2
 \approx \biggl(\frac{\alpha_s(E)}{\pi}\biggr)^2,  \label{b1rate}
\end{equation}  
where $f_{b1}\approx f_{\pi}$ has been used. Eq. (\ref{b1rate}) is 
the prediction of short-distance dominance.  
 We should keep in mind that 
the prediction on $b_1^{\pm}\pi^{\mp}$ involves the same uncertainties 
as we have mentioned for $a_2^{\pm}\pi^{\mp}$ at the end of the 
preceding section. Nonetheless, it is safe 
to state with Eqs. (\ref{a+}), (\ref{a-}), and (\ref{b1rate}) that 
$b_1^-\pi^+$ may be produced at the level of $10^{-6}$ or so if
the perturbative QCD picture is correct for final-state interactions, but 
that the $b_1^+\pi^-$ decay should not be seen at any level ($< 10^{-9}$)
in that case.

\section{Helicity conservation and spin structure}

The predictions in the preceding sections lead us to conclude that
the $B^0$-$\overline{B}^0$ interference can be observed in the 
$a_0^{\pm}\pi^{\mp}$ or the $b_1^{\pm}\pi^{\mp}$ channel only if 
a very strong enhancement occurs by long-distance interactions in 
an otherwise suppressed mode. In such a case, the classification of
amplitudes by the tree and the penguin interaction becomes a bit blurry.

One powerful test exists for the short-distance dominance. For a 
two-body decay where both final mesons have spin, $J$ and $J'(\geq J)$, 
we have $2J+1$ independent amplitudes. If short-distance interactions 
dominate, the decay into the zero-helicity state should dominate over 
all other helicity states in the charmless $B$ decay. Let us explain 
it briefly since this is a robust prediction of the Standard Model 
and provides a simple experimental test of short-distance dominance 
independent of the rate measurement. 

For the tree operators in which all quark fields are left-chiral,
one of the final mesons must be formed with the fast $q_L$ and 
$\overline{q_L}$ without involving the spectator. Since these quarks 
remember their helicities throughout interactions with hard gluons, 
the resulting $q_L\overline{q_L}$ meson state must be in
helicity zero ($h=0$) in the approximation of ignoring the quark mass
and the higher configuration such as $q\overline{q}g$ and 
$q\overline{q}q\overline{q}$. Then the other meson that picks up
the spectator is forced to have $h=0$ by overall angular momentum 
conservation. This argument applies to the penguin operators 
${\cal O}_{3,4}$ too. In the case of ${\cal O}_{5,6}$, the argument is 
the same when $\overline{q}_R$ and $q_R$ from ${\cal O}_{5,6}$ form 
one meson. If instead $\overline{d_L}$ ($\overline{s_L}$) and $q_R$ 
form a meson, 
this meson would have $h=+1$, and the other meson, being formed 
with $\overline{q_R}$ and the spectator, can only be in $h=0$ or $-1$.  
The overall angular conservation therefore forbids this decay. 
No matter which interaction causes a decay, therefore, the dominant 
final helicity state is zero.\footnote{
Actually, it is sufficient to prove this selection rule for the 
fundamental weak interaction ($\sim{\cal O}_2$) since all other decay 
operators are generated from it through hard gluon-loop corrections.}  
Production of $h=\pm 1$ is allowed to the extent of the nonzero quark 
masses and of the transverse motion of quarks inside a meson, 
{\em i.e.,} the meson mass.

  A remark is in order concerning the quark mass effect in the $a_0$ 
production. The $a_0^+$ meson is produced by the operator 
$\overline{u_R}d_L$ of ${\cal O}_{5,6}$ which produces $h=+1$, while 
the spinless $a_0^+$ cannot have nonzero helicity. This means 
that, if we ignore the quark mass, production of $a_0^+$ is forbidden. 
The $a_0^+$ production occurs through the small $h=0$ component of 
$O(\sqrt{m_q^2+{\bf k}_{\perp}^2}/E_p)$ that is contained in 
$\overline{u_R}d_L$. This remark applies to $b_1^+$ too. 
That is the reason why the production amplitudes of
$a_0^+$ and $b_1^+$ are suppressed by one power of $1/m_B$.

When the mass and ${\bf k}_{\perp}$ corrections are included, the $h=+1$ 
amplitude is generated with $O(m/E)$ for the $B(\overline{b}q)$ decay 
while the $h=-1$ amplitude is generated only with $O(m^2/E^2)$. 
The reason is as follows: The spectator quark can be either in $h= +
\frac{1}{2}$ or $-\frac{1}{2}$ with a 50/50 chance. Therefore the
$h=+1$ decay amplitude can be realized with the small opposite 
helicity component of a single fast $q_L$ (out of $\overline{q_L}
q_L\overline{q_L}$ from ${\cal O}_{1\sim 4}$) or $\overline{q_R}$ 
(out of $\overline{d_L}q_R\overline{q_R}$ from ${\cal O}_{5,6}$). 
On the other hand the $h=-1$ amplitude needs small components of 
two $\overline{q_L}$'s (from ${\cal O}_{1\sim 4}$) or $q_R$ and 
$\overline{q_L}$ (from ${\cal O}_{5,6}$). Since each small component 
costs $m_T/E_p\sim 2m_T/m_B$, we obtain the following hierarchy of 
helicity suppression for the helicity amplitudes ($A_h \equiv
\langle M({\bf p},h) M'(-{\bf p},h)|{\cal H}_{int}|B(\overline{b}q)
\rangle$):
\begin{eqnarray}
        A_{+1}/A_0 &\approx& m/m_B, \nonumber \\ 
        A_{-1}/A_0 &\approx& mm'/m_B^2, \label{helicity}
\end{eqnarray}
where $m$ and $m'$ are the meson masses. In the first line,
$m$ stands for the mass of the meson formed with the fast 
$q$ and $\overline{q}$ emitted from the current, not with the spectator.
Because the small opposite helicity of this meson forces the
other meson to have $h=+1$ state. 

The $h=0$ dominance can be easily tested in experiment by
measuring the angular distribution of the decay products of either 
of final mesons\cite{Dighe} although distinguishing between $h=+1$ 
and $-1$ is much harder\cite{Chiang}.
Actually, the zero-helicity dominance holds for some processes 
involving a charm quark such as $B^0\to D^{*-}\rho^+$ for which 
the tree interaction completely dominates. Since $\rho^+$ is 
produced from the $V-A$ current of the tree interaction, its helicity 
must be zero. Therefore the $h=0$ dominance should hold to 
$O(m_{\rho}/m_B)$ in this case. The percentage of the $h=0$ branching 
was found in experiment as $\Gamma_L= (93\pm 5 \pm 5)\%$ for
$\overline{B}^0\to D^{*+}\rho^-$\cite{Drho}.  The selection 
rule holds in a modified form for $B\to J/\psi K^*$\cite{Suzuki}.  
A significant correction may arise to the $h=0$ dominance 
because of the large $c$-quark mass. However, the $K^*$ can 
only be in $h=0$ or $+1$ for $m_{K^*}\ll m_B$ and therefore
$|A_{+1}|\gg |A_{-1}|$ should hold well. 
Experiment\cite{BaBar2} is consistent with $|A_{+1}|\gg |A_{-1}|$, 
but cannot distinguish it from $|A_{-1}|\gg |A_{+1}|$ until
more a sophisticated cascade decay measurement is made. The 
zero-helicity dominance rule is yet to be tested in the charmless 
$B$ decay.

The zero-helicity dominance can be found in the early literature.
Ali {\em et al}\cite{Ali} computed the $B\to VV$ decay in the 
factorization of the tree interaction with the $U(6)\times U(6)$ 
form factors which happen to incorporate helicity conservation. 
One can read off the suppression pattern of Eq. (\ref{helicity})
in their results, as Chen {\em et al.,}\cite{Chen} recently pointed 
out. K\"{o}rner and Goldstein\cite{Korner} discussed 
it for charm decays. If the final-state interaction is included 
through hadron rescattering, the helicity selection rule breaks 
down in general. We can show, for instance, that
if one computes $B\to\pi\pi\to\rho\rho$ with $\omega$-exchange for
$\pi\pi\to\rho\rho$, the final $\rho$'s are polarized
dominantly in $h=\pm 1$. This illustrates that the $h=0$ 
dominance easily breaks down by long-distance final-state interactions.
If on the other hand one computes the $\pi\pi\to\rho\rho$ rescattering
with $\pi$-exchange, the $\rho$'s are polarized mostly in $h=0$.
Whether short-distances dominate or not should be determined
eventually by experiment.

Test of the zero-helicity dominance has one clear advantage over the
branching fraction measurement. There is a limitation in accuracy
even in the short-distance calculation when one evaluates the magnitudes 
and phases of amplitudes and sum them up. In the hadron picture, 
it is nearly impossible to compute final-state interactions and
predict decay amplitudes reliably. Therefore the branching fraction 
measurement alone will be inconclusive in determining how much 
long-distance contributions exist in a given decay process. If we 
combine it with the zero-helicity dominance test, we can be more 
confident with our conclusions since the rule should hold to all 
orders of QCD as long as they are of short distances

\section{Summary}

If short-distance physics dominates in final states, it will be
nearly impossible to obtain the weak angle from the time-dependent
decay  $B^0/\overline{B}^0\to a_0^{\pm}\pi^{\mp}(\rho^{\mp})$ or  
$b_1^{\pm}\pi^{\mp}(\rho^{\mp})$. While the flavor-tagged measurement 
will tell us about short-distance dominance in these decays, more 
a general and simple test is to examine zero-helicity dominance in 
the charmless $B$ decay with spins.

\section{Acknowledgment}
This work was supported in part by the Director, Office of Science, 
Office of High Energy and Nuclear Physics, Division of High Energy 
Physics, of the U.S.  Department of Energy under contract 
DE-AC03-76SF00098 and in part by the National Science Foundation 
under grant PHY-0098840.

\end{document}